\documentclass[aps,prb,twocolumn,superscriptaddress,showpacs]{revtex4-1}
\usepackage{hyperref}
\usepackage{amsmath}
\usepackage{graphicx}

\begin{document}

\title{Complete phase diagram for three-band Hubbard model with orbital degeneracy lifted by crystal field splitting}
\author{Li Huang}
\affiliation{ Beijing National Laboratory for Condensed Matter Physics, 
              and Institute of Physics, 
              Chinese Academy of Sciences, 
              Beijing 100190, 
              China }
\affiliation{ National Key Laboratory for Surface Physics and Chemistry, 
              P.O. Box 718-35, 
              Mianyang 621907, 
              Sichuan, 
              China }

\author{Liang Du}
\affiliation{ Beijing National Laboratory for Condensed Matter Physics, 
              and Institute of Physics, 
              Chinese Academy of Sciences, 
              Beijing 100190, 
              China }

\author{Xi Dai}
\affiliation{ Beijing National Laboratory for Condensed Matter Physics, 
              and Institute of Physics, 
              Chinese Academy of Sciences, 
              Beijing 100190, 
              China }

\date{\today}

\begin{abstract}
Motivated by the unexplored complexity of the phase diagrams for multi-orbital Hubbard models, 
a three-band Hubbard model at integer fillings ($N=4$) with orbital degeneracy lifted partially
by crystal field splitting is analyzed systematically in this work. By using single site dynamical 
mean-field theory and rotationally invariant Gutzwiller approximation, we have computed the full 
phase diagram with Coulomb interaction strength $U$ and crystal field splitting $\Delta$. We find 
a large region in the phase diagram, where an orbital selective Mott phase will be stabilized by 
the positive crystal field lifting the orbital degeneracy. Further analysis indicates that the 
Hund's rule coupling is essential for the orbital selective Mott phase and the transition toward 
this phase is accompanied by a high-spin to low-spin transition. Such a model may be relevant for 
the recently discovered pnictides and ruthenates correlated materials.
\end{abstract}


\maketitle

\section{introduction}
\label{sec:intro}

The Mott-Hubbard metal-insulator transition (MIT) has been a subject of great interest for
decades.\cite{imada:1039} Most of the attentions before this century have been focused on 
the one band case only, because most of the qualitative features of MIT have already been 
captured by single band Hubbard model, as shown by numerous studies using the single site 
dynamical mean-field theory (DMFT).\cite{georges:13} Whereas in realistic materials most of 
the Mott transitions involve more than one band and thus exhibit multi-orbital features.\cite{tokura:462}
Multi-orbital extension of the Hubbard model allows more realistic description of MIT and 
other strongly correlated physics, which contains, in general, much richer phase diagrams, 
and exotic physical phenomena. For instance, the redistribution of electrons among different 
orbitals leads to new scenarios such as orbital ordering,\cite{chan:235114} high-spin 
to low-spin (HS-LS) transition\cite{werner:126405} or orbital selective Mott transition 
(OSMT)\cite{anisimov:191} etc.

Mott transitions in multi-orbital models have been studied within the framework of DMFT for 
more than ten years.\cite{georges:13,kotliar:865} Previous studies show\cite{florens:205102} 
that simple increasing the band degeneracy only changes the critical interaction strength 
$U_c$, but will not change the fundamental features of Mott transition, where all the degenerate 
bands undergo Mott transition simultaneously under the increment of interaction strength. 
Recent studies of realistic materials have focused interest on the interplay between the MIT 
and orbital degeneracy.\cite{anisimov:191,shorikov:195101,kunes:198,craco:064425} A very 
fundamental question raised in this field is, how the multi-orbital systems response to the 
breaking down of the orbital degeneracy. In such a system, it is possible that the Mott 
transitions in different orbitals happen separately, which is the so-called OSMT and has 
been suggested firstly by Anisimov \emph{et al.}\cite{anisimov:191} in the study of Ca$_{2-x}$Sr$_{x}$RuO$_{4}$.

After the concept of OSMT has been proposed, lots of research interests  have been attracted.
\cite{koga:216402,koga:045128,liebsch:226401,liebsch:165103,knecht:081103,arita:201102,dongen:2006,
ferrero:205126,werner:115119,jakobi:115109,neupane:097001,medici:126401,de'medici:205124}
The early DMFT studies on this problem are focused on the two-band Hubbard model with half
filling, which is the simplest system that may have OSMT when the bandwidths of the two 
bands are different. The DMFT calculations from different groups with exact diagonalization 
(ED)\cite{koga:216402,koga:045128} and Hirsch-Fye quantum Monte Carlo (HFQMC)\cite{liebsch:226401,
liebsch:165103,arita:201102,knecht:081103,dongen:2006} as impurity solvers converge to two 
essential conclusions: (1) The OSMT in two-band Hubbard model is mainly induced by the bandwidth 
difference which breaks the degeneracy between the two bands and the crystal field splitting 
plays a minor role here. (2) The emergence of OSMT is very sensitive to the feature of the 
local interaction. It is very easy to occur OSMT when the local interaction is rotationally 
invariant, while very difficult to happen when the local interaction breaks the rotational 
invariance.

The OSMT in the three-band Hubbard model, which is more relevant to the realistic situation of
Ca$_{2-x}$Sr$_{x}$RuO$_{4}$,\cite{anisimov:191,neupane:097001} is not a trivial generalization 
of the two-band model. In two-band systems, the OSMT can only happen in half filling case, while 
in three-band systems it can happen when the occupation numbers are 2, 3 or 4. When the total 
occupation number is 3, the three-band system is half filling and actually the situation is very 
similar with the two-band case. The most interesting case is when the occupation number is 2 
or 4, which can be transformed between each other by particle-hole symmetry. Recently de' Medici 
\emph{et al}.\cite{medici:126401} proposed a new mechanism for OSMT, which happens in 
three-band systems with filling factor being 2 or 4. In this new scenario of OSMT, the driving 
force is not the difference of bandwidth but the crystal field splitting lifting the band 
degeneracy. And then Kita \emph{et al.}\cite{kita:195130} investigated how the orbital level 
splitting and Ising-type Hund's rule coupling affect the Mott transition in the case of two electrons 
per site. Their results reveal that the critical interaction strength separating a metallic 
phase and two kinds of insulating phases shows a non-monotonic behavior as a function of the 
level splitting. They suggest that this behavior is characteristic for 1/3 filling, in comparison 
with the preceding results for different fillings and for two-band models. We note that the 
three-band system is very popular in transition metal compounds.\cite{imada:1039,tokura:462} 
Provided the Fermi energy falls into the $t_{2g}$ bands, the tetragonal distortion will further 
split the $t_{2g}$ bands into non-degenerate $a_{1g}$ band and two-fold degenerate $e'_{g}$ bands. 
When the occupation number is 4, the appropriate crystal field will redistribute the four 
electrons into $a_{1g}$ band and $e'_{g}$ bands as 1 and 3 respectively (i.e (3,1) configuration). 
Thus if we neglect the correlation between $a_{1g}$ and $e'_{g}$ sub-systems, the $a_{1g}$ band 
becomes a one-band system at half filling and $e'_{g}$ bands become a two-band system at quarter 
filling, which will lead to an orbital selective Mott phase (OSMP) at equal bandwidth because it is 
much easier to get Mott insulator in $a_{1g}$ band. The mean-field phase diagram of this model 
determined by slave-spin method\cite{medici:126401,de'medici:205124} contains quite a large 
region for OSMP, in which the $a_{1g}$ band has already become Mott insulator while the $e'_{g}$ 
bands are still metallic.

The OSMT driven by the band degeneracy lifting is quite a robust phenomena determined by 
the interplay between crystal symmetry and correlation effects. In reference [\onlinecite{medici:126401}], 
the phase diagram of the three-band Hubbard model is mainly calculated by the slave-spin 
method, which is qualitatively correct but not accurate enough. In the present paper, we 
study systematically this $t_{2g}$ Hubbard model with crystal field splitting 
by two more accurate methods: the DMFT method combined with state-of-the-art hybridization 
expansion continuous time quantum Monte Carlo (CTQMC) impurity solver\cite{werner:155107,
werner:076405,gull:349,gull:20111078} and newly developed rotationally invariant Gutzwiller 
approximation (RIGA) method.\cite{weber:6896,bunemann:236404,deng:075114,lanata:1108.0180} 
The metal-insulator phase diagram, band specific quasiparticle weight $Z_a$ and occupation 
number $n_a$ are computed by the both methods with respect to crystal field 
splitting $\Delta$ and Coulomb interaction strength $U$. Based on these results, we mainly 
discuss three important aspects of OSMT in this system: (1) The crucial role of Hund's 
rule coupling; (2) The redistribution of four electrons among $a_{1g}$ and $e'_{g}$ bands; 
(3) The relationship between OSMT and HS-LS spin state crossover.

The rest of this paper is organized as follows: In Sec.\ref{sec:model} the
three-band Hubbard model treated in this work is specified. In Sec.\ref{subsec:rot} the 
main results of this paper, $U-\Delta$ phase diagrams for rotationally invariant interaction 
and SU($N$) density-density interaction are presented and compared with each other. In 
Sec.\ref{subsec:red}, the redistribution of electrons among different orbitals and its
relationship with the OSMT 
are discussed in detail. And the accompanying HS-LS spin state crossovers are discussed 
in Sec.\ref{subsec:hl}. Section \ref{sec:conclusion} serves as a conclusion.

\section{model}
\label{sec:model}

We consider the three-band Hubbard model defined by
\begin{equation}
\label{eq:ham}
H= -\sum_{i,j,a,\sigma}t_{ij} c^{\dagger}_{ia\sigma} c_{ja\sigma} + \sum_i H^{i}_{loc},
\end{equation}
where $c^{(\dagger)}_{ia\sigma}$ is an annihilation (creation) operator of an electron
with spin $\sigma$ (=$\uparrow$, $\downarrow$) and orbital $a$ (=1, 2, 3) at the $i$th
site, $t_{ij}$ is the hopping integral between site $i$ and site $j$. The local part of
Hamiltonian $H^{i}_{loc}$ can be defined as follows (for the sake of simplicity, the site 
index $i$ has been ignored):
\begin{equation}
\label{eq:loc}
\begin{split}
H_{loc} = &- \sum_{a,\sigma}(\mu - \Delta_a) n_{a,\sigma} + \sum_{a} Un_{a,\uparrow}n_{a,\downarrow} \\
          &+ \sum_{a > b,\sigma}[U'n_{a,\sigma}n_{b,-\sigma} + (U'-J)n_{a,\sigma}n_{b,\sigma}] \\
          &- \sum_{a < b} J (d^{\dagger}_{a,\downarrow}d^{\dagger}_{b,\uparrow}d_{b,\downarrow}d_{a,\uparrow} 
           + d^{\dagger}_{b,\uparrow}d^{\dagger}_{b,\downarrow}d_{a,\uparrow}d_{a,\downarrow} + h.c.).
\end{split}
\end{equation}
Here $n_{a,\sigma} = c^{\dagger}_{a\sigma}c_{a\sigma}$ is the number operator, $\mu$ is the 
chemical potential, and $\Delta_a$ is the energy level for orbital $a$. In the interaction 
terms, $U(U')$ is the intra-orbital (inter-orbital) Coulomb interaction, and $J$ is the Hund's rule 
coupling. The constrained condition $U = U' + 2J$ is imposed as usual, which is valid for atomic 
like local orbitals. The above interaction terms include both the spin-flip and pair-hopping 
terms, and thus are rotational invariant in the spin space. In order to elucidate the Mott MIT 
in the case of four electrons per site, the chemical potential $\mu$ is adjusted dynamically in 
the simulations to fix the electron filling per site as $N = 4$. In the present paper, we focus 
on the $t_{2g}$-like bands under the tetragonal crystal field, which are split into non-degenerate 
$a_{1g}$ band and double degenerate $e'_{g}$ bands. Therefore the on-site energy level is assumed 
to be $\Delta_1 \neq \Delta_2 = \Delta_3$ and their difference is defined as $\Delta = \Delta_1 - \Delta_2 $.

This lattice model (see Eq.(\ref{eq:ham})) can be solved in the framework of single site DMFT\cite{georges:13,kotliar:865}, 
which neglects the momentum dependence of the self-energy and reduces the original lattice problem 
to the self-consistent solution of an effective impurity model. In this paper, a semicircular density 
of states with half bandwidth $D=1$ is used, which corresponds to the infinite coordination Bethe 
lattice. All the orbitals have equal bandwidth and the energy unit is set to be $D$. To solve the 
effective impurity model, the hybridization expansion CTQMC impurity solver\cite{werner:155107,werner:076405,
gull:349,gull:20111078} is adopted. This method allows us to access the strong interaction regime 
down to very low temperature. In our calculations, the system temperature is set to be $T=0.01$ 
(corresponding to inverse temperature $\beta = 100$) unless otherwise stated. In each DMFT iterations, 
typically $4 \times 10^{8}$ QMC samplings have been performed to reach sufficient numerical accuracy.

In the present works, we also use the RIGA method\cite{weber:6896,bunemann:236404,deng:075114,
lanata:1108.0180} to crosscheck the results obtained by DMFT+CTQMC method. In contrast to the finite 
temperature DMFT+CTQMC calculations, only the zero temperature physical quantities can be obtained 
by using the RIGA method. Although the RIGA method can not access the dynamical properties of 
correlated systems, it provides a very fast and economic way to calculate the ground state 
properties, for examples, the total energy and occupation numbers. And it can be used as a 
good complementary method to the DMFT approach. As will be discussed in detail in the following 
sections, the phase diagrams and other physical properties obtained by both methods are in very 
good agreements. The implementation details about our RIGA method have been presented in 
reference [\onlinecite{lanata:1108.0180}].

\section{Results and discussion}
\label{sec:results}

\subsection{$U-\Delta$ phase diagram}
\label{subsec:rot}

\begin{figure}
\centering
\includegraphics[scale=0.6]{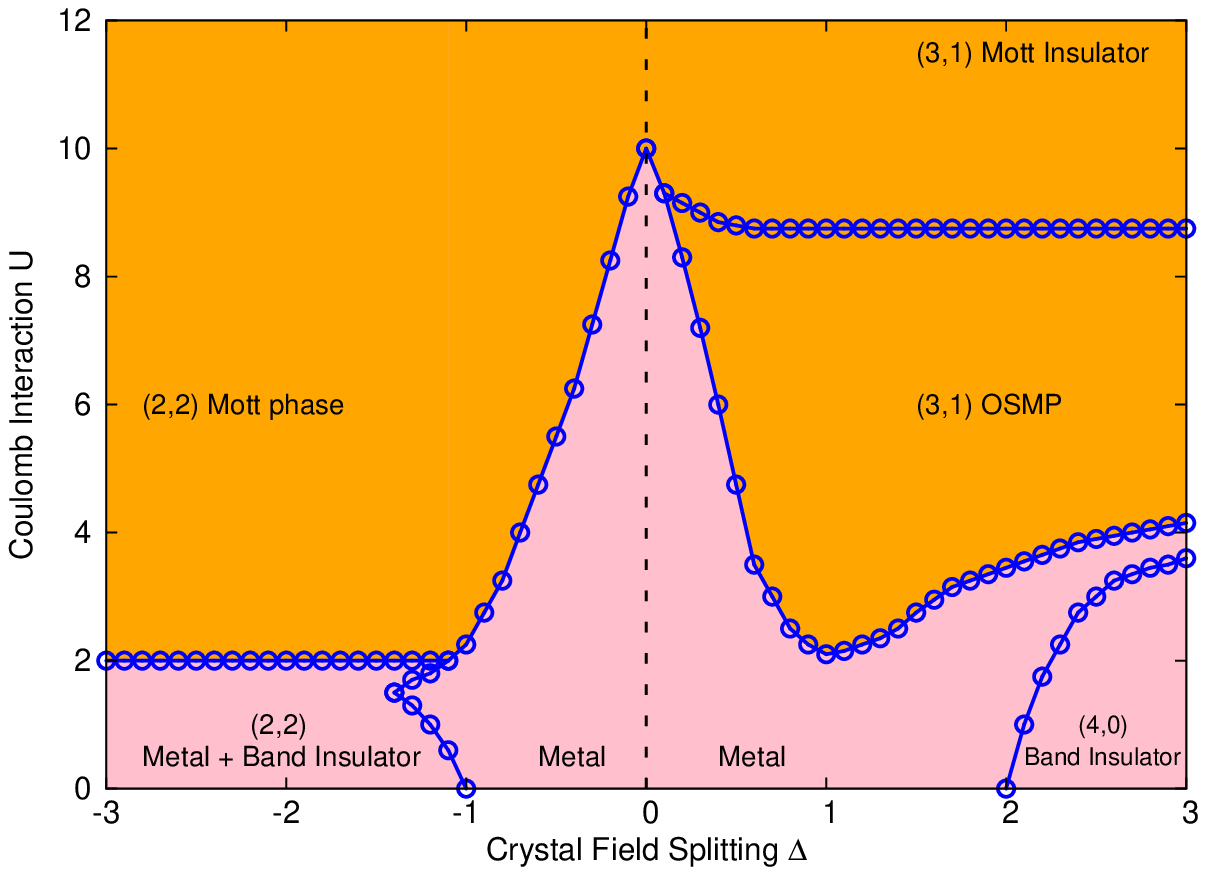}
\includegraphics[scale=0.6]{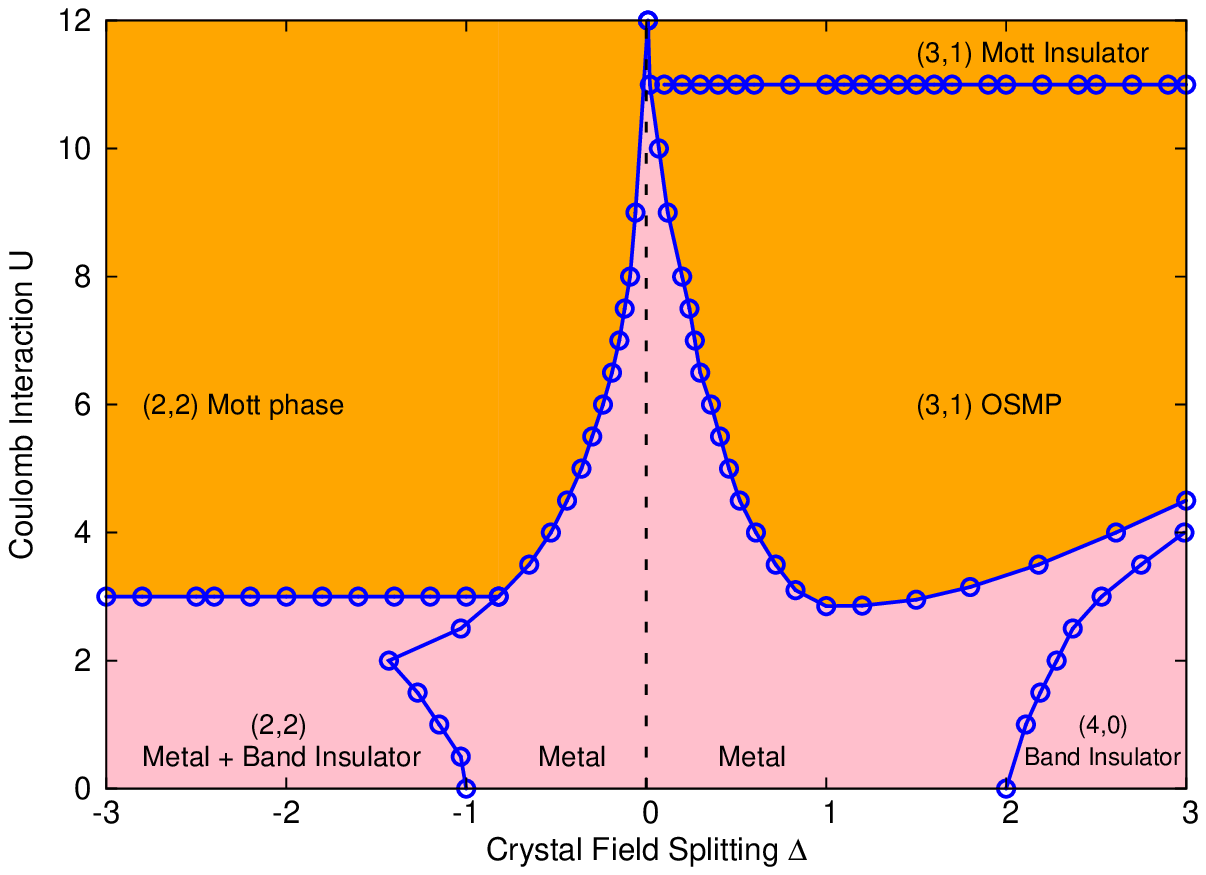}
\caption{(Color online) Calculated phase diagram of three-band Hubbard model with rotationally 
invariant interactions in the plane of Coulomb interaction $U$ ($J = U/4$) and crystal field splitting 
$\Delta$ ($\Delta = \Delta_1 - \Delta_2$). Upper panel: Calculated by DMFT+CTQMC method at 
finite temperature $T=0.01$. Lower panel: Calculated by RIGA method at zero temperature. In 
all the calculations, the chemical potential $\mu$ is adjusted dynamically to fulfill the 
total occupation number condition ($N=4$). The label ``(2,2) Metal + Band Insulator" means the 
two-fold degenerate bands ($e'_{g}$ states) are metallic while the non-degenerate band 
($a_{1g}$ state) is insulating, and (2,2) means corresponding orbital occupancies. All 
the other labels have similar explanations. The pink 
zone means LS state and the orange zone means HS state respectively.\label{ph1}}
\end{figure}

\begin{figure}
\centering
\includegraphics[scale=0.6]{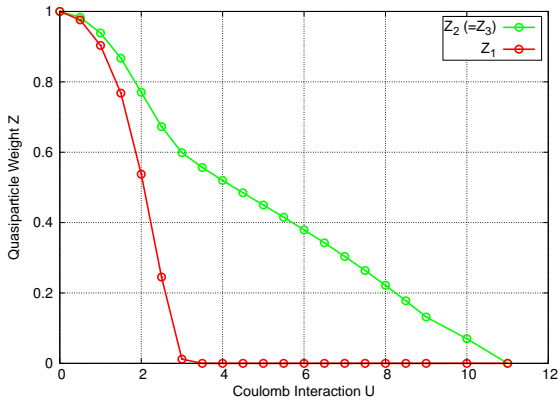}
\caption{(Color online) Quasiparticle weights $Z_a$ as a function of Coulomb interaction $U$ 
for selected crystal field splitting $\Delta = 1.0$. The calculations are done by the RIGA method 
at zero temperature. \label{xz}}
\end{figure}

\begin{figure}
\includegraphics[scale=0.6]{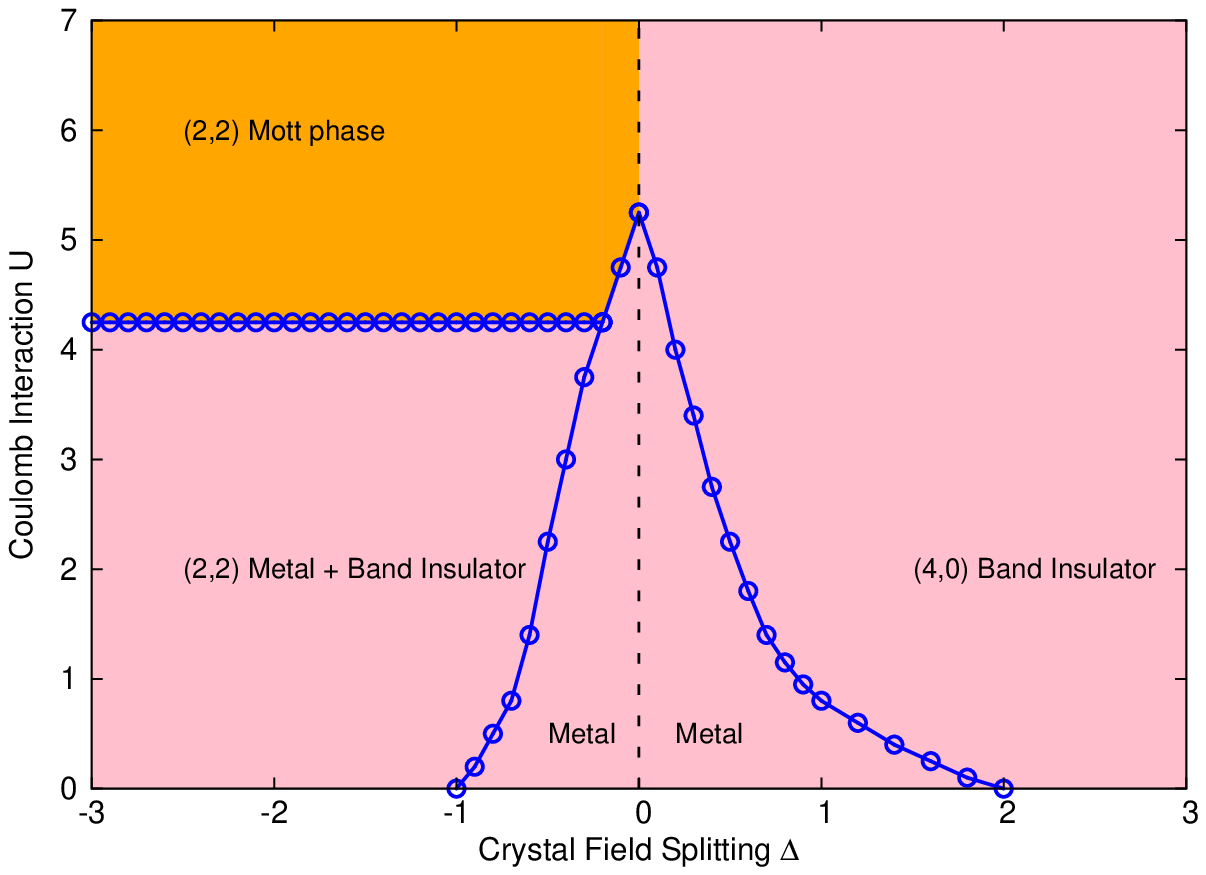}
\includegraphics[scale=0.6]{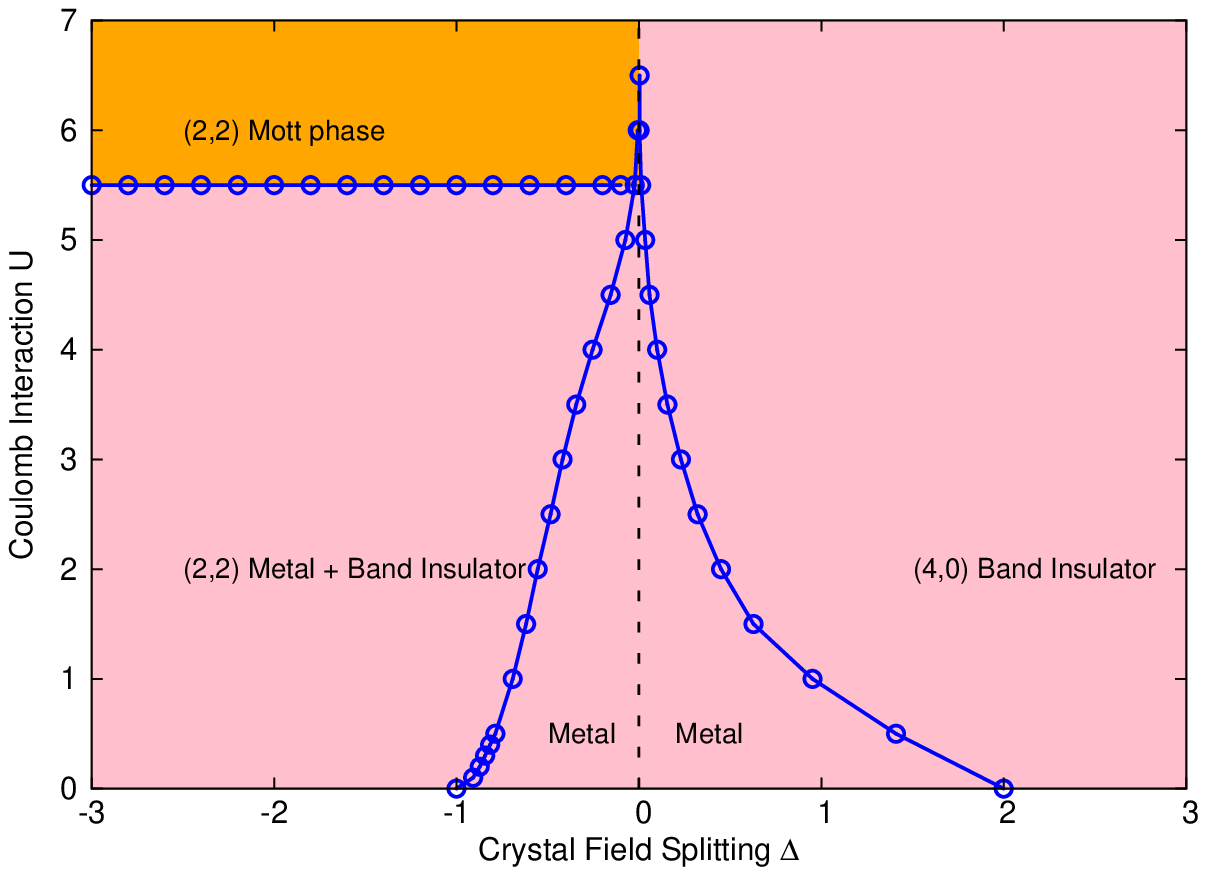}
\caption{(Color online) Calculated phase diagram of three-band Hubbard model with 
SU($N$) density-density interaction in the plane of Coulomb interaction $U$ ($J = 0.0$) 
and crystal field splitting $\Delta$ ($\Delta = \Delta_1 - 
\Delta_2$). Upper panel: Calculated by DMFT+CTQMC method at finite temperature $T=0.01$; Lower 
panel: Calculated by RIGA method at zero temperature. In the calculations, the chemical 
potential $\mu$ is adjusted dynamically to fulfill the total occupation number condition 
($N=4$). The label ``(2,2) Metal + Band Insulator" means the two-fold degenerate bands 
($e'_{g}$ states) are metallic while the non-degenerate band ($a_{1g}$ state) is insulating, 
and (2,2) means corresponding orbital occupancies. 
All the other labels have similar explanations. The pink zone means LS state and the orange 
zone means HS state respectively.\label{ph3}}
\end{figure}

To map out the metal-insulator phase diagram we have computed systematically the dependence 
of charge density, quasiparticle weight, and local magnetic moment as the function of crystal field 
splitting for various Coulomb interaction strengths. In this subsection, we will focus firstly on 
the $U-\Delta$ phase diagrams with both non-zero and zeroed Hund's rule coupling $J$. We will show in the following 
that the Hund's rule coupling $J$ is extremely crucial for the appearance of OSMP in the three-band 
systems with four electrons.

When the spin-flip and pair-hopping interaction terms are taken into full considerations, the Hamiltonian 
for three-band Hubbard model is rotationally invariant in the spin space, and the phase 
diagram contains four different phases, which are metal, band insulator, OSMP and Mott 
insulator phases, respectively. The obtained $U-\Delta$ phase diagrams for $J=U/4$ case are 
illustrated in Fig.\ref{ph1}. The upper panel is the results obtained by 
DMFT+CTQMC method, and the lower panel is obtained by RIGA method. From the above figure, we 
can easily see that the consistency between both two methods is quite excellent, except the 
phase boundary obtained by RIGA method is a bit higher, which is similar with the situations 
found in one and two-band models.\cite{deng:075114,lanata:1108.0180} Comparing with the similar 
phase diagrams obtained by slave-spin method in the literatures,\cite{medici:126401,de'medici:205124} 
there are two important differences. Firstly, the critical $U_c$ for Mott transition at 
$\Delta=0$ is much lower in the phase diagram obtained by DMFT+CTQMC method. Secondly, in 
the phase diagram obtained by both DMFT+CTQMC and RIGA methods, the phase boundary between 
the metal phase and band insulator phase depends on the interaction strength $U$ monotonically, 
however in the phase diagram obtained by slave-spin method, it decreases firstly and then 
increases.\cite{medici:126401} This difference is mainly due to the over simplified treatment 
of Hund's rule coupling terms in the slave-spin method.\cite{de'medici:205124}

The general shape of the phase diagram can be easily understood by considering two 
limiting cases: (1) For $\Delta=0$, the model reduces to a fully degenerate three-band 
Hubbard model with total filling $N=4$, which undergoes a Mott transition around
$U = 10.0$; (2) For the uncorrelated limit ($U=0$), when $\Delta > 0$, 
a simple transition from metal to band insulator can be observed at $\Delta=2.0$ with 
the fully occupied $e'_{g}$ bands and empty $a_{1g}$ band. Another transition can also 
be seen on the negative side at $\Delta=-1.0$, after which the non-degenerate $a_{1g}$ 
band is fully occupied band insulator and the double degenerate $e'_{g}$ bands are 
metallic at half filling.

When both the Coulomb interaction $U$ and crystal field $\Delta$ are non-zero, the phase 
diagram is quite complicated and very different for $\Delta > 0$ and $\Delta < 0$. Now 
let's make a further discussion about this phase diagram. If the crystal field splitting 
$\Delta > 0$, the isolate $a_{1g}$ band is lifted up and the double degenerate $e'_{g}$ bands 
are pushed down. This part of phase diagram can be divided vertically into two regions: 
(1) When $ 2.0 > \Delta > 0.0$, the system undergoes successive transitions from metal to 
Mott insulator through an OSMP as the Coulomb interaction strength increases; (2) When 
$\Delta > 2.0$, the system undergoes a different type of successive transitions from 
band insulator to Mott insulator through metal and OSMP phases in sequence. On the other 
hand, if the crystal field splitting $\Delta < 0$, i.e, the $a_{1g}$ band is lower than 
the two-fold degenerate $e'_{g}$ bands, this part of phase diagram can be divided into 
three different regions: (1) For $-1.0 < \Delta < 0.0$, with weak Coulomb interaction both 
bands have fractional filling and are metallic. As the increment of Coulomb interaction 
$U$, the system undergoes a transition from a fully metallic phase to an insulator phase, 
after which the occupation numbers for both $a_{1g}$ and $e'_{g}$ bands are two. Since the 
$a_{1g}$ band has no degeneracy, it is completely occupied and becomes band insulator in 
this situation. While at the same time the $e'_{g}$ bands with double degeneracy become Mott 
insulator with half filling. It is called $(2,2)$ Mott insulator phase throughout this paper. 
(2) When $-1.4 < \Delta < -1.0$, with weak Coulomb interaction the $a_{1g}$ band is already 
fully occupied and the double degenerate $e'_{g}$ bands are half-filled and metallic. With 
the increment of Coulomb interaction $U$, the system first becomes fully metallic phase with 
fractional filling factors for all the bands, and then goes back to the original phase after 
a ``reentrance" transition. Finally it becomes the $(2,2)$ Mott insulator phase for $U>2.0$. 
(3) When $\Delta < -1.4$, the $a_{1g}$ band remains fully occupied regardless of the interaction 
strength and the system reduces to a two-band system with half filling, which undergoes a 
typical Mott MIT around $U=2.0$ as determined by DMFT+CTQMC method.

We note that the effect of Coulomb interaction $U$ have two important effects for multi-orbital 
systems. One is to reduce the quasiparticle weight, the other one is to redistribute the electrons 
among different bands. The interplay between these two effects determines the main structure of the 
above phase diagram. Next we will focus on the first effect, and the second effect is less emphasized 
before and will be discussed in detail in the following subsections.

The band specific quasiparticle weights $Z_a$ as a function of Coulomb interaction $U$ 
are shown in Fig.\ref{xz}. The crystal field splitting is fixed to $\Delta = 1.0$. 
For simplicity, only the results obtained by RIGA method, which are consistent with 
those obtained by DMFT+CTQMC method, are displayed in this figure. It is apparent that 
the quasiparticle weights $Z_a$ decrease monotonously from 1.0 to 0.0 when Coulomb interaction strength increases. 
As $U < 3.0$, the $Z_a$ are large than 0.1, and both the $e'_{g}$ and $a_{1g}$ bands are
metallic. As $ U > 3.0 $, $Z_1$ turns to zero while $Z_2$($=Z_{3}$) is still finite. It means
that the $a_{1g}$ band undergoes a Mott MIT and turns into insulator at $U = 3.0$, and at 
the same time the $e'_{g}$ bands keep metallic, which is the so-called (3,1) OSMP in the 
phase diagram. When Coulomb interaction strength continues to increase ($U > 11.0$ for the 
RIGA method and $U > 9.0$ for the DMFT+CTQMC method), the $e'_{g}$ bands
undergo another Mott MIT and the system becomes the (3,1) Mott insulator phase finally, 
then all $Z_a$ will approach zero.
 
The Hund's rule coupling $J$ has enormous influence on the metal-insulator phase diagram for 
multi-orbital Hubbard model.\cite{medici:256401,kita:195130} It is noted that if we neglect 
the spin-flip and pair-hopping terms, only keep the Ising-type Hund's rule coupling term 
$J_z$ ($=J$), we can obtain very similar phase diagram, albeit the phase boundary is shifted 
downwards slightly. It is apparent that the Ising-type Hund's rule coupling term $J_z$ is the 
key requirement for OSMT and such a rich phase diagram. In order to reveal the underlying 
physics more clearly we also did calculations for the phase diagram of similar three-band 
Hubbard model without Hund's rule coupling terms ($J=0$). Thus we can compare the $U-\Delta$ 
phase diagrams for the case with finite $J$ (see Fig.\ref{ph1}) and SU($N$) $J=0$ case (see 
Fig.\ref{ph3}). Obviously the latter is much simpler and has no OSMP in the whole phase diagram.

The calculated $U-\Delta$ phase diagram for $J=0$ case is illustrated in Fig.\ref{ph3}. The upper 
panel is the calculated results obtained by DMFT+CTQMC method, and the lower panel is obtained by 
RIGA method. These two methods give almost identical results again. Firstly we concentrate on the 
non-interaction case ($U = 0$). When the crystal field splitting is positive ($\Delta > 0$), an 
obvious transition can be seen at $\Delta = 2.0$ after which the system becomes band insulator. 
On the other hand, when crystal field splitting is negative ($\Delta < 0$), similar MIT can be 
found at $\Delta =-1.0$. When $\Delta > 0$, the phase diagram only consists of metal and band 
insulator phases. The lower left region is metallic, and the upper right region is band 
insulator phase. The OSMP is disappeared completely. When $\Delta < 0$, the phase diagram 
contains metal, band insulator, and Mott insulator phases, but the characteristic ``tip" which 
can be clearly seen at $U \sim 1.0$ and $\Delta \sim -1.2 $ in Fig.\ref{ph1} vanishes, and the 
phase boundary between metal plus band insulator and (2,2) Mott insulator is shifted upward 
significantly. Thus, it is concluded that the Hund's rule coupling terms $J$ has 
played a key role on the phase diagram and finite $J_z$ is the minimal requirement to drive an OSMT.

\subsection{Redistribution of electrons}
\label{subsec:red}

\begin{figure}
\centering
\includegraphics[scale=0.6]{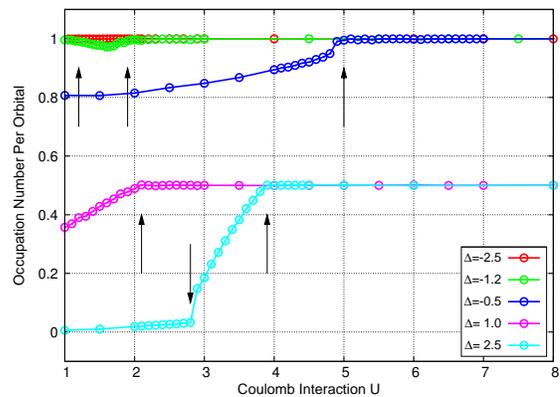}
\caption{(Color online) Orbital filling as a function of Coulomb interaction $U$ for selected 
crystal field splitting $\Delta$ values (rotationally invariant interaction case). The results 
are calculated by DMFT+CTQMC method at finite temperature $T=0.01$. In this figure, only the 
electron densities of non-degenerate band ($a_{1g}$ band) are shown. The arrows 
correspond to phase transition points.\label{xn}}
\end{figure}

\begin{figure}
\centering
\includegraphics[scale=0.6]{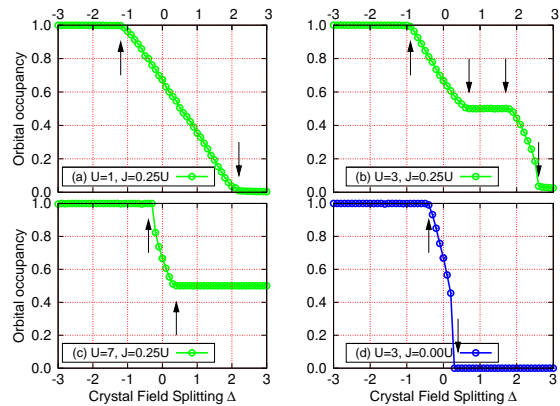}
\caption{(Color online) Orbital filling as a function of crystal field splitting $\Delta$ for 
selected Coulomb interaction $U$ values. (a) $U = 1.0$; (b) $U = 3.0$; (c) $U = 7.0$; (d) 
$U = 3.0$. In (a)-(c), the interaction term is rotationally invariant with non-zero spin-flip 
and pair-hopping terms, whereas in (d) the interaction term is SU($N$) scheme with $J = 0$. All 
the calculations are done by DMFT+CTQMC method at finite temperature $T=0.01$. In this figure, 
only the electron densities of non-degenerate $a_{1g}$ band are shown. The arrows correspond 
to phase transition points.\label{hn}}
\end{figure}

\begin{figure}
\includegraphics[scale=0.6]{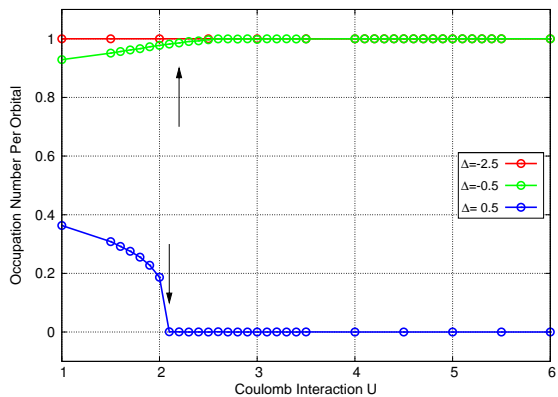}
\caption{(Color online) Orbital filling as a function of Coulomb interaction $U$ for selected 
crystal field splitting values (SU($N$) $J=0$ case). All the calculations are done by DMFT+CTQMC 
method at finite temperature $T=0.01$. In this figure, only the electron densities of 
non-degenerate $a_{1g}$ band are shown. The arrows correspond to phase transition points. \label{zn}}
\end{figure}

In this subsection, we will discuss the second important effect of Coulomb interaction $U$ on
multi-orbital systems with broken symmetry: the redistribution of electrons among different orbitals. In order to 
understand the intriguing physics contained in the $U-\Delta$ phase diagrams more clearly, 
we further plot the occupation numbers of non-degenerate $a_{1g}$ band as functions of Coulomb 
interaction $U$ and crystal field splitting $\Delta$ in Fig.\ref{xn}$\sim$\ref{zn}, respectively.

In Fig.\ref{xn}, the evolution of the orbital occupancy with the Coulomb interaction strength is 
shown. The vertical arrows denote phase transition points. In this figure, only the orbital 
filling of the non-degenerate band ($a_{1g}$ band) is plotted. For negative crystal field, 
$a_{1g}$ band is much lower in energy, and the effect of Coulomb interaction depends on the value 
of the crystal field $\Delta$ in the following way. For $0>\Delta>-1.0$, the effect of correlation 
effect is to transfer electrons from $e'_{g}$ bands to the $a_{1g}$ band until it is fully occupied 
and becomes band insulator, as shown by the blue curve in Fig.\ref{xn}. For $-1.0>\Delta>-1.4$, 
the $a_{1g}$ band is already fully occupied in the non-interacting case and the effect of the 
correlation is non-monotonic. As the increment of the repulsive interaction $U$ and Hund's rule 
coupling $J$, the occupation number of $a_{1g}$ band first drops due to the Hund's rule coupling 
and then returns back to be fully occupied. We note that the interesting non-monotonic behavior 
of the occupation is the consequence of the interplay between Hund's rule coupling $J$, which 
favors even distribution of the electrons, and the Coulomb repulsive interaction $U$, which intends 
to increase the occupation difference between orbitals. Therefore, this behavior disappears when 
the Hund's rule coupling has been set to be zero, as shown in figure \ref{zn}. While for 
$\Delta<-1.4$, the charge distribution will not be affected by the correlation effect and the 
occupation number of $a_{1g}$ band keeps constant as the increment of $U$.

While the situation is very different for $\Delta > 0 $, in this case the energy level of 
$a_{1g}$ band is higher and the occupation number is less than $0.5$ in the non-interacting case. 
As we can see from Fig.\ref{xn}, the increment of the interaction strength will pump the 
electrons from $e'_{g}$ bands to $a_{1g}$ band again until it reaches half filling and 
becomes Mott insulator. When the crystal field strength $\Delta$ is smaller than $2.0$ all 
the bands are metallic, and the system becomes OSMP when the $a_{1g}$ band reaches half 
filling. While for $\Delta$ is larger than $2.0$, the system starts from a typical band insulator
with fully occupied $e'_{g}$ and empty $a_{1g}$ bands. With the increment of Coulomb 
interaction, it first becomes metallic when the $a_{1g}$ band get partially populated 
and finally goes into the OSMP. 

In Fig.\ref{hn} (a)-(c), we plot the occupancy of $a_{1g}$ band under fixed Coulomb interaction 
strength $U$ as a function of crystal field splitting $\Delta$. For weak Coulomb interaction 
($U=1.0$), when the crystal field splitting is increased from very negative to very positive value,
two plateaus can be found in the occupancy of $a_{1g}$ band, which correspond to fully occupied and
empty situations respectively. The occupation numbers decrease smoothly between the two plateaus 
corresponding to the metallic phase. For intermediate interaction strength with non-zero
Hund's rule coupling $J$, there is additional plateau with occupancy being half filling, which 
corresponds to the OSMT phase which is completely absent when $J=0$.

The redistribution of the electrons among the three orbitals is the key point for the OSMT in 
this system and can be understood by the effect of Hund's rule coupling, which favors the HS 
state with the (3,1) configuration. As for SU($N$) $J=0$ case, the redistribution of electrons 
is much more easy to be understood. As shown in Fig.\ref{zn}, the correlation effect induced 
by the Coulomb interaction increases the occupation of $a_{1g}$ band for $\Delta < 0$ case 
and decreases it for $\Delta > 0$ case, which is consistent with the results obtained by 
Hartree-Fock mean-field method.

\subsection{HS-LS transition}
\label{subsec:hl}

\begin{figure}
\centering
\includegraphics[scale=0.6]{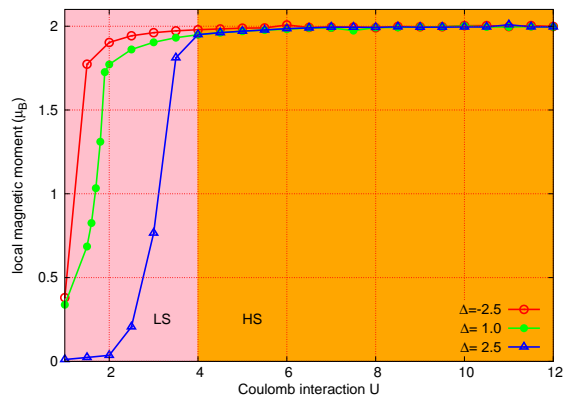}
\caption{(Color online) The calculated effective local magnetic moment $\sqrt{\langle S^{2}_{z}\rangle}$
for three-band Hubbard model with rotationally invariant interaction terms. The calculations are done
by DMFT+CTQMC method at finite temperature $T=0.01$. In this figure, only the calculated results for
representative crystal field splitting values ($\Delta=$-2.5,\ 1.0, and 2.5) are presented. The pink 
region denotes LS phase, while orange region denotes HS phase.\label{sz}}
\end{figure}

In this subsection, we focus on the magnetic properties of the three-band model 
during the phase transitions which have been reported rarely in the references. In 
Fig.\ref{sz} we plot the evolution of the mean instantaneous moment, which is defined 
as $\sqrt{\langle S^2_z\rangle}$, as the function of Coulomb interaction strength $U$ 
with selected crystal field splitting $\Delta$. We find that although the overall behavior 
looks quite similar for positive and negative crystal field splitting, the underlying 
physics are very different for the two cases.

For the negative crystal field splitting $\Delta < 0$, with the increment of Coulomb 
interaction strength and Hund's rule coupling, the system is approaching the $(2,2)$ 
configuration (two electrons in both $a_{1g}$ and $e'_{g}$ bands). The HS state can be 
reached smoothly when the two electrons in the $e'_{g}$ bands fall into the triplet 
state due to the Hund's rule coupling. At the mean while, the Mott transition happens 
when the spin state gets close to the HS state.

Nevertheless, the situation on the positive crystal field splitting side $\Delta > 0$ 
is very different. Under weak Coulomb interaction, the occupation numbers of the system 
are more close to the $(4,0)$ configuration (all the four electrons are in $e'_{g}$ bands). 
With the increment of Coulomb interaction strength and Hund's rule coupling, when 
the crystal field splitting is still not strong enough to drive the system into the 
band insulator, the system undergoes two sequential transition. The first transition 
is more related to the charge transfer from the $e'_{g}$ bands to $a_{1g}$ band and 
the system changes from $(4,0)$ to $(3,1)$ configuration afterwards. This transition 
comes from the competition between the crystal field splitting, which favors the $(4,0)$ 
configuration, and the Hund's rule coupling, which favors the $(3,1)$ configuration. 
After the first transition, which manifests itself in Fig.\ref{sz} as a clear kink in 
the $\sqrt{\langle S^{2}_{z}\rangle} - U$ curve, the $a_{1g}$ band is half filling 
and the $e'_{g}$ bands are only quarter filling. Since the system with half filling 
is always much closer to a Mott transition, the $a_{1g}$ band becomes Mott insulator 
first after the first transition, while the $e'_{g}$ bands still remain metallic. A 
second transition happens by further increasing the Coulomb interaction together with 
the Hund's rule coupling, after which the system becomes Mott insulator for all the 
bands with the spin state being very close to a pure HS state.

\section{concluding remarks}
\label{sec:conclusion}

We have studied the Mott transition in the three-band Hubbard model with orbital degeneracy lifting 
by crystal field splitting. By using the DMFT+CTQMC method and RIGA method, we have investigated how
the orbital level splitting and the Hund's rule coupling affect the Mott transition in a system with 
four electrons per site. We obtain the following conclusions. 

Firstly, the Hund's rule coupling $J$ (more specifically, the Ising-type Hund's rule coupling $J_z$) 
is the minimal requirement to induce OSMT and stabilize OSMP. In the phase diagram for three-band 
Hubbard model with rotationally invariant interactions, the OSMP covers a wide parameter range. 
While in the phase diagram for three-band Hubbard model without Hund's rule coupling terms, though 
the Mott MIT occurs when the Coulomb interaction strength reaches critical values, but the OSMP 
is absent. 

Secondly, the interplay between crystal field splitting and Hund's rule coupling terms induces the 
redistribution of electrons among the three bands and leads to the complex phase diagrams. For 
examples, the (4,0) electronic configuration is corresponding to a band insulator phase, (3,1) 
configuration leads to OSMP or Mott insulator phase, and (2,2) configuration leads to 
metal plus band insulator or Mott insulator plus band insulator. 

Thirdly, the appearance of the OSMP in this system is always accompanied by a HS-LS spin state 
crossover. In the OSMP, the electron distribution keeps in (3,1) configuration, which strongly 
favors the HS state and lowers the Hund's rule energy. When the Coulomb interaction is reduced, the OSMP 
collapses to metal phase accompanied by a HS-LS transition.

Finally, three-band models arise in a number of other physically important contexts, including
doped C$_{60}$, ruthenates, etc. Extending our results to models with more realistic
density of states is of high priority for future research.

\begin{acknowledgments}
We acknowledge financial support from the National Science Foundation ͑of China and that
from the 973 program of China under Contract No.2007CB925000 and No.2011CBA00108. The 
DMFT+CTQMC calculations have been performed on the SHENTENG7000 at Supercomputing Center 
of Chinese Academy of Sciences (SCCAS).
\end{acknowledgments}

\bibliography{osmt}

\begin{thebibliography}{34}%
\makeatletter
\providecommand \@ifxundefined [1]{%
 \@ifx{#1\undefined}
}%
\providecommand \@ifnum [1]{%
 \ifnum #1\expandafter \@firstoftwo
 \else \expandafter \@secondoftwo
 \fi
}%
\providecommand \@ifx [1]{%
 \ifx #1\expandafter \@firstoftwo
 \else \expandafter \@secondoftwo
 \fi
}%
\providecommand \natexlab [1]{#1}%
\providecommand \enquote  [1]{``#1''}%
\providecommand \bibnamefont  [1]{#1}%
\providecommand \bibfnamefont [1]{#1}%
\providecommand \citenamefont [1]{#1}%
\providecommand \href@noop [0]{\@secondoftwo}%
\providecommand \href [0]{\begingroup \@sanitize@url \@href}%
\providecommand \@href[1]{\@@startlink{#1}\@@href}%
\providecommand \@@href[1]{\endgroup#1\@@endlink}%
\providecommand \@sanitize@url [0]{\catcode `\\12\catcode `\$12\catcode
  `\&12\catcode `\#12\catcode `\^12\catcode `\_12\catcode `\%12\relax}%
\providecommand \@@startlink[1]{}%
\providecommand \@@endlink[0]{}%
\providecommand \url  [0]{\begingroup\@sanitize@url \@url }%
\providecommand \@url [1]{\endgroup\@href {#1}{\urlprefix }}%
\providecommand \urlprefix  [0]{URL }%
\providecommand \Eprint [0]{\href }%
\providecommand \doibase [0]{http://dx.doi.org/}%
\providecommand \selectlanguage [0]{\@gobble}%
\providecommand \bibinfo  [0]{\@secondoftwo}%
\providecommand \bibfield  [0]{\@secondoftwo}%
\providecommand \translation [1]{[#1]}%
\providecommand \BibitemOpen [0]{}%
\providecommand \bibitemStop [0]{}%
\providecommand \bibitemNoStop [0]{.\EOS\space}%
\providecommand \EOS [0]{\spacefactor3000\relax}%
\providecommand \BibitemShut  [1]{\csname bibitem#1\endcsname}%
\let\auto@bib@innerbib\@empty
\bibitem [{\citenamefont {Imada}\ \emph {et~al.}(1998)\citenamefont {Imada},
  \citenamefont {Fujimori},\ and\ \citenamefont {Tokura}}]{imada:1039}%
  \BibitemOpen
  \bibfield  {author} {\bibinfo {author} {\bibfnamefont {M.}~\bibnamefont
  {Imada}}, \bibinfo {author} {\bibfnamefont {A.}~\bibnamefont {Fujimori}}, \
  and\ \bibinfo {author} {\bibfnamefont {Y.}~\bibnamefont {Tokura}},\
  }\href@noop {} {\bibfield  {journal} {\bibinfo  {journal} {Rev. Mod. Phys.}\
  }\textbf {\bibinfo {volume} {70}},\ \bibinfo {pages} {1039} (\bibinfo {year}
  {1998})}\BibitemShut {NoStop}%
\bibitem [{\citenamefont {Georges}\ \emph {et~al.}(1996)\citenamefont
  {Georges}, \citenamefont {Kotliar}, \citenamefont {Krauth},\ and\
  \citenamefont {Rozenberg}}]{georges:13}%
  \BibitemOpen
  \bibfield  {author} {\bibinfo {author} {\bibfnamefont {A.}~\bibnamefont
  {Georges}}, \bibinfo {author} {\bibfnamefont {G.}~\bibnamefont {Kotliar}},
  \bibinfo {author} {\bibfnamefont {W.}~\bibnamefont {Krauth}}, \ and\ \bibinfo
  {author} {\bibfnamefont {M.~J.}\ \bibnamefont {Rozenberg}},\ }\href@noop {}
  {\bibfield  {journal} {\bibinfo  {journal} {Rev. Mod. Phys.}\ }\textbf
  {\bibinfo {volume} {68}},\ \bibinfo {pages} {13} (\bibinfo {year}
  {1996})}\BibitemShut {NoStop}%
\bibitem [{\citenamefont {Tokura}\ and\ \citenamefont
  {Nagaosa}(2000)}]{tokura:462}%
  \BibitemOpen
  \bibfield  {author} {\bibinfo {author} {\bibfnamefont {Y.}~\bibnamefont
  {Tokura}}\ and\ \bibinfo {author} {\bibfnamefont {N.}~\bibnamefont
  {Nagaosa}},\ }\href@noop {} {\bibfield  {journal} {\bibinfo  {journal}
  {Science}\ }\textbf {\bibinfo {volume} {288}},\ \bibinfo {pages} {462}
  (\bibinfo {year} {2000})}\BibitemShut {NoStop}%
\bibitem [{\citenamefont {Chan}\ \emph {et~al.}(2009)\citenamefont {Chan},
  \citenamefont {Werner},\ and\ \citenamefont {Millis}}]{chan:235114}%
  \BibitemOpen
  \bibfield  {author} {\bibinfo {author} {\bibfnamefont {C.-K.}\ \bibnamefont
  {Chan}}, \bibinfo {author} {\bibfnamefont {P.}~\bibnamefont {Werner}}, \ and\
  \bibinfo {author} {\bibfnamefont {A.~J.}\ \bibnamefont {Millis}},\
  }\href@noop {} {\bibfield  {journal} {\bibinfo  {journal} {Phys. Rev. B}\
  }\textbf {\bibinfo {volume} {80}},\ \bibinfo {pages} {235114} (\bibinfo
  {year} {2009})}\BibitemShut {NoStop}%
\bibitem [{\citenamefont {Werner}\ and\ \citenamefont
  {Millis}(2007)}]{werner:126405}%
  \BibitemOpen
  \bibfield  {author} {\bibinfo {author} {\bibfnamefont {P.}~\bibnamefont
  {Werner}}\ and\ \bibinfo {author} {\bibfnamefont {A.~J.}\ \bibnamefont
  {Millis}},\ }\href@noop {} {\bibfield  {journal} {\bibinfo  {journal} {Phys.
  Rev. Lett.}\ }\textbf {\bibinfo {volume} {99}},\ \bibinfo {pages} {126405}
  (\bibinfo {year} {2007})}\BibitemShut {NoStop}%
\bibitem [{\citenamefont {Anisimov}\ \emph {et~al.}(2002)\citenamefont
  {Anisimov}, \citenamefont {Nekrasov}, \citenamefont {Kondakov}, \citenamefont
  {Rice},\ and\ \citenamefont {Sigrist}}]{anisimov:191}%
  \BibitemOpen
  \bibfield  {author} {\bibinfo {author} {\bibfnamefont {V.~I.}\ \bibnamefont
  {Anisimov}}, \bibinfo {author} {\bibfnamefont {I.~A.}\ \bibnamefont
  {Nekrasov}}, \bibinfo {author} {\bibfnamefont {D.~E.}\ \bibnamefont
  {Kondakov}}, \bibinfo {author} {\bibfnamefont {T.~M.}\ \bibnamefont {Rice}},
  \ and\ \bibinfo {author} {\bibfnamefont {M.}~\bibnamefont {Sigrist}},\
  }\href@noop {} {\bibfield  {journal} {\bibinfo  {journal} {Eur. Phys. J. B}\
  }\textbf {\bibinfo {volume} {25}},\ \bibinfo {pages} {191} (\bibinfo {year}
  {2002})}\BibitemShut {NoStop}%
\bibitem [{\citenamefont {Kotliar}\ \emph {et~al.}(2006)\citenamefont
  {Kotliar}, \citenamefont {Savrasov}, \citenamefont {Haule}, \citenamefont
  {Oudovenko}, \citenamefont {Parcollet},\ and\ \citenamefont
  {Marianetti}}]{kotliar:865}%
  \BibitemOpen
  \bibfield  {author} {\bibinfo {author} {\bibfnamefont {G.}~\bibnamefont
  {Kotliar}}, \bibinfo {author} {\bibfnamefont {S.~Y.}\ \bibnamefont
  {Savrasov}}, \bibinfo {author} {\bibfnamefont {K.}~\bibnamefont {Haule}},
  \bibinfo {author} {\bibfnamefont {V.~S.}\ \bibnamefont {Oudovenko}}, \bibinfo
  {author} {\bibfnamefont {O.}~\bibnamefont {Parcollet}}, \ and\ \bibinfo
  {author} {\bibfnamefont {C.~A.}\ \bibnamefont {Marianetti}},\ }\href@noop {}
  {\bibfield  {journal} {\bibinfo  {journal} {Rev. Mod. Phys.}\ }\textbf
  {\bibinfo {volume} {78}},\ \bibinfo {pages} {865} (\bibinfo {year}
  {2006})}\BibitemShut {NoStop}%
\bibitem [{\citenamefont {Florens}\ \emph {et~al.}(2002)\citenamefont
  {Florens}, \citenamefont {Georges}, \citenamefont {Kotliar},\ and\
  \citenamefont {Parcollet}}]{florens:205102}%
  \BibitemOpen
  \bibfield  {author} {\bibinfo {author} {\bibfnamefont {S.}~\bibnamefont
  {Florens}}, \bibinfo {author} {\bibfnamefont {A.}~\bibnamefont {Georges}},
  \bibinfo {author} {\bibfnamefont {G.}~\bibnamefont {Kotliar}}, \ and\
  \bibinfo {author} {\bibfnamefont {O.}~\bibnamefont {Parcollet}},\ }\href@noop
  {} {\bibfield  {journal} {\bibinfo  {journal} {Phys. Rev. B}\ }\textbf
  {\bibinfo {volume} {66}},\ \bibinfo {pages} {205102} (\bibinfo {year}
  {2002})}\BibitemShut {NoStop}%
\bibitem [{\citenamefont {Shorikov}\ \emph {et~al.}(2010)\citenamefont
  {Shorikov}, \citenamefont {Pchelkina}, \citenamefont {Anisimov},
  \citenamefont {Skornyakov},\ and\ \citenamefont {Korotin}}]{shorikov:195101}%
  \BibitemOpen
  \bibfield  {author} {\bibinfo {author} {\bibfnamefont {A.~O.}\ \bibnamefont
  {Shorikov}}, \bibinfo {author} {\bibfnamefont {Z.~V.}\ \bibnamefont
  {Pchelkina}}, \bibinfo {author} {\bibfnamefont {V.~I.}\ \bibnamefont
  {Anisimov}}, \bibinfo {author} {\bibfnamefont {S.~L.}\ \bibnamefont
  {Skornyakov}}, \ and\ \bibinfo {author} {\bibfnamefont {M.~A.}\ \bibnamefont
  {Korotin}},\ }\href@noop {} {\bibfield  {journal} {\bibinfo  {journal} {Phys.
  Rev. B}\ }\textbf {\bibinfo {volume} {82}},\ \bibinfo {pages} {195101}
  (\bibinfo {year} {2010})}\BibitemShut {NoStop}%
\bibitem [{\citenamefont {Kunes}\ \emph {et~al.}(2008)\citenamefont {Kunes},
  \citenamefont {Lukoyanov}, \citenamefont {Anisimov}, \citenamefont
  {Scalettar},\ and\ \citenamefont {Pickett}}]{kunes:198}%
  \BibitemOpen
  \bibfield  {author} {\bibinfo {author} {\bibfnamefont {J.}~\bibnamefont
  {Kunes}}, \bibinfo {author} {\bibfnamefont {A.~V.}\ \bibnamefont
  {Lukoyanov}}, \bibinfo {author} {\bibfnamefont {V.~I.}\ \bibnamefont
  {Anisimov}}, \bibinfo {author} {\bibfnamefont {R.~T.}\ \bibnamefont
  {Scalettar}}, \ and\ \bibinfo {author} {\bibfnamefont {W.~E.}\ \bibnamefont
  {Pickett}},\ }\href@noop {} {\bibfield  {journal} {\bibinfo  {journal}
  {Nature Materials}\ }\textbf {\bibinfo {volume} {7}},\ \bibinfo {pages} {198}
  (\bibinfo {year} {2008})}\BibitemShut {NoStop}%
\bibitem [{\citenamefont {Craco}\ \emph {et~al.}(2006)\citenamefont {Craco},
  \citenamefont {Laad},\ and\ \citenamefont
  {M\"{u}ller-Hartmann}}]{craco:064425}%
  \BibitemOpen
  \bibfield  {author} {\bibinfo {author} {\bibfnamefont {L.}~\bibnamefont
  {Craco}}, \bibinfo {author} {\bibfnamefont {M.~S.}\ \bibnamefont {Laad}}, \
  and\ \bibinfo {author} {\bibfnamefont {E.}~\bibnamefont
  {M\"{u}ller-Hartmann}},\ }\href@noop {} {\bibfield  {journal} {\bibinfo
  {journal} {Phys. Rev. B}\ }\textbf {\bibinfo {volume} {74}},\ \bibinfo
  {pages} {064425} (\bibinfo {year} {2006})}\BibitemShut {NoStop}%
\bibitem [{\citenamefont {Koga}\ \emph {et~al.}(2004)\citenamefont {Koga},
  \citenamefont {Kawakami}, \citenamefont {Rice},\ and\ \citenamefont
  {Sigrist}}]{koga:216402}%
  \BibitemOpen
  \bibfield  {author} {\bibinfo {author} {\bibfnamefont {A.}~\bibnamefont
  {Koga}}, \bibinfo {author} {\bibfnamefont {N.}~\bibnamefont {Kawakami}},
  \bibinfo {author} {\bibfnamefont {T.~M.}\ \bibnamefont {Rice}}, \ and\
  \bibinfo {author} {\bibfnamefont {M.}~\bibnamefont {Sigrist}},\ }\href@noop
  {} {\bibfield  {journal} {\bibinfo  {journal} {Phys. Rev. Lett.}\ }\textbf
  {\bibinfo {volume} {92}},\ \bibinfo {pages} {216402} (\bibinfo {year}
  {2004})}\BibitemShut {NoStop}%
\bibitem [{\citenamefont {Koga}\ \emph {et~al.}(2005)\citenamefont {Koga},
  \citenamefont {Kawakami}, \citenamefont {Rice},\ and\ \citenamefont
  {Sigrist}}]{koga:045128}%
  \BibitemOpen
  \bibfield  {author} {\bibinfo {author} {\bibfnamefont {A.}~\bibnamefont
  {Koga}}, \bibinfo {author} {\bibfnamefont {N.}~\bibnamefont {Kawakami}},
  \bibinfo {author} {\bibfnamefont {T.~M.}\ \bibnamefont {Rice}}, \ and\
  \bibinfo {author} {\bibfnamefont {M.}~\bibnamefont {Sigrist}},\ }\href@noop
  {} {\bibfield  {journal} {\bibinfo  {journal} {Phys. Rev. B}\ }\textbf
  {\bibinfo {volume} {72}},\ \bibinfo {pages} {045128} (\bibinfo {year}
  {2005})}\BibitemShut {NoStop}%
\bibitem [{\citenamefont {Liebsch}(2003)}]{liebsch:226401}%
  \BibitemOpen
  \bibfield  {author} {\bibinfo {author} {\bibfnamefont {A.}~\bibnamefont
  {Liebsch}},\ }\href@noop {} {\bibfield  {journal} {\bibinfo  {journal} {Phys.
  Rev. Lett.}\ }\textbf {\bibinfo {volume} {91}},\ \bibinfo {pages} {226401}
  (\bibinfo {year} {2003})}\BibitemShut {NoStop}%
\bibitem [{\citenamefont {Liebsch}(2004)}]{liebsch:165103}%
  \BibitemOpen
  \bibfield  {author} {\bibinfo {author} {\bibfnamefont {A.}~\bibnamefont
  {Liebsch}},\ }\href@noop {} {\bibfield  {journal} {\bibinfo  {journal} {Phys.
  Rev. B}\ }\textbf {\bibinfo {volume} {70}},\ \bibinfo {pages} {165103}
  (\bibinfo {year} {2004})}\BibitemShut {NoStop}%
\bibitem [{\citenamefont {Knecht}\ \emph {et~al.}(2005)\citenamefont {Knecht},
  \citenamefont {Bl\"umer},\ and\ \citenamefont {van Dongen}}]{knecht:081103}%
  \BibitemOpen
  \bibfield  {author} {\bibinfo {author} {\bibfnamefont {C.}~\bibnamefont
  {Knecht}}, \bibinfo {author} {\bibfnamefont {N.}~\bibnamefont {Bl\"umer}}, \
  and\ \bibinfo {author} {\bibfnamefont {P.~G.~J.}\ \bibnamefont {van
  Dongen}},\ }\href@noop {} {\bibfield  {journal} {\bibinfo  {journal} {Phys.
  Rev. B}\ }\textbf {\bibinfo {volume} {72}},\ \bibinfo {pages} {081103}
  (\bibinfo {year} {2005})}\BibitemShut {NoStop}%
\bibitem [{\citenamefont {Arita}\ and\ \citenamefont
  {Held}(2005)}]{arita:201102}%
  \BibitemOpen
  \bibfield  {author} {\bibinfo {author} {\bibfnamefont {R.}~\bibnamefont
  {Arita}}\ and\ \bibinfo {author} {\bibfnamefont {K.}~\bibnamefont {Held}},\
  }\href@noop {} {\bibfield  {journal} {\bibinfo  {journal} {Phys. Rev. B}\
  }\textbf {\bibinfo {volume} {72}},\ \bibinfo {pages} {201102} (\bibinfo
  {year} {2005})}\BibitemShut {NoStop}%
\bibitem [{\citenamefont {Dongen}\ \emph {et~al.}(2006)\citenamefont {Dongen},
  \citenamefont {Knecht},\ and\ \citenamefont {Bl{\"u}mer}}]{dongen:2006}%
  \BibitemOpen
  \bibfield  {author} {\bibinfo {author} {\bibfnamefont {P.~G. J.~v.}\
  \bibnamefont {Dongen}}, \bibinfo {author} {\bibfnamefont {C.}~\bibnamefont
  {Knecht}}, \ and\ \bibinfo {author} {\bibfnamefont {N.}~\bibnamefont
  {Bl{\"u}mer}},\ }\href@noop {} {\bibfield  {journal} {\bibinfo  {journal}
  {Phys. Status Solidi (B)}\ }\textbf {\bibinfo {volume} {243}},\ \bibinfo
  {pages} {116} (\bibinfo {year} {2006})}\BibitemShut {NoStop}%
\bibitem [{\citenamefont {Ferrero}\ \emph {et~al.}(2005)\citenamefont
  {Ferrero}, \citenamefont {Becca}, \citenamefont {Fabrizio},\ and\
  \citenamefont {Capone}}]{ferrero:205126}%
  \BibitemOpen
  \bibfield  {author} {\bibinfo {author} {\bibfnamefont {M.}~\bibnamefont
  {Ferrero}}, \bibinfo {author} {\bibfnamefont {F.}~\bibnamefont {Becca}},
  \bibinfo {author} {\bibfnamefont {M.}~\bibnamefont {Fabrizio}}, \ and\
  \bibinfo {author} {\bibfnamefont {M.}~\bibnamefont {Capone}},\ }\href@noop {}
  {\bibfield  {journal} {\bibinfo  {journal} {Phys. Rev. B}\ }\textbf {\bibinfo
  {volume} {72}},\ \bibinfo {pages} {205126} (\bibinfo {year}
  {2005})}\BibitemShut {NoStop}%
\bibitem [{\citenamefont {Werner}\ \emph {et~al.}(2009)\citenamefont {Werner},
  \citenamefont {Gull},\ and\ \citenamefont {Millis}}]{werner:115119}%
  \BibitemOpen
  \bibfield  {author} {\bibinfo {author} {\bibfnamefont {P.}~\bibnamefont
  {Werner}}, \bibinfo {author} {\bibfnamefont {E.}~\bibnamefont {Gull}}, \ and\
  \bibinfo {author} {\bibfnamefont {A.~J.}\ \bibnamefont {Millis}},\
  }\href@noop {} {\bibfield  {journal} {\bibinfo  {journal} {Phys. Rev. B}\
  }\textbf {\bibinfo {volume} {79}},\ \bibinfo {pages} {115119} (\bibinfo
  {year} {2009})}\BibitemShut {NoStop}%
\bibitem [{\citenamefont {Jakobi}\ \emph {et~al.}(2009)\citenamefont {Jakobi},
  \citenamefont {Bl\"umer},\ and\ \citenamefont {van Dongen}}]{jakobi:115109}%
  \BibitemOpen
  \bibfield  {author} {\bibinfo {author} {\bibfnamefont {E.}~\bibnamefont
  {Jakobi}}, \bibinfo {author} {\bibfnamefont {N.}~\bibnamefont {Bl\"umer}}, \
  and\ \bibinfo {author} {\bibfnamefont {P.}~\bibnamefont {van Dongen}},\
  }\href@noop {} {\bibfield  {journal} {\bibinfo  {journal} {Phys. Rev. B}\
  }\textbf {\bibinfo {volume} {80}},\ \bibinfo {pages} {115109} (\bibinfo
  {year} {2009})}\BibitemShut {NoStop}%
\bibitem [{\citenamefont {Neupane}\ \emph {et~al.}(2009)\citenamefont
  {Neupane}, \citenamefont {Richard}, \citenamefont {Pan}, \citenamefont {Xu},
  \citenamefont {Jin}, \citenamefont {Mandrus}, \citenamefont {Dai},
  \citenamefont {Fang}, \citenamefont {Wang},\ and\ \citenamefont
  {Ding}}]{neupane:097001}%
  \BibitemOpen
  \bibfield  {author} {\bibinfo {author} {\bibfnamefont {M.}~\bibnamefont
  {Neupane}}, \bibinfo {author} {\bibfnamefont {P.}~\bibnamefont {Richard}},
  \bibinfo {author} {\bibfnamefont {Z.-H.}\ \bibnamefont {Pan}}, \bibinfo
  {author} {\bibfnamefont {Y.-M.}\ \bibnamefont {Xu}}, \bibinfo {author}
  {\bibfnamefont {R.}~\bibnamefont {Jin}}, \bibinfo {author} {\bibfnamefont
  {D.}~\bibnamefont {Mandrus}}, \bibinfo {author} {\bibfnamefont
  {X.}~\bibnamefont {Dai}}, \bibinfo {author} {\bibfnamefont {Z.}~\bibnamefont
  {Fang}}, \bibinfo {author} {\bibfnamefont {Z.}~\bibnamefont {Wang}}, \ and\
  \bibinfo {author} {\bibfnamefont {H.}~\bibnamefont {Ding}},\ }\href@noop {}
  {\bibfield  {journal} {\bibinfo  {journal} {Phys. Rev. Lett.}\ }\textbf
  {\bibinfo {volume} {103}},\ \bibinfo {pages} {097001} (\bibinfo {year}
  {2009})}\BibitemShut {NoStop}%
\bibitem [{\citenamefont {de' Medici}\ \emph {et~al.}(2009)\citenamefont {de'
  Medici}, \citenamefont {Hassan}, \citenamefont {Capone},\ and\ \citenamefont
  {Dai}}]{medici:126401}%
  \BibitemOpen
  \bibfield  {author} {\bibinfo {author} {\bibfnamefont {L.}~\bibnamefont {de'
  Medici}}, \bibinfo {author} {\bibfnamefont {S.~R.}\ \bibnamefont {Hassan}},
  \bibinfo {author} {\bibfnamefont {M.}~\bibnamefont {Capone}}, \ and\ \bibinfo
  {author} {\bibfnamefont {X.}~\bibnamefont {Dai}},\ }\href@noop {} {\bibfield
  {journal} {\bibinfo  {journal} {Phys. Rev. Lett.}\ }\textbf {\bibinfo
  {volume} {102}},\ \bibinfo {pages} {126401} (\bibinfo {year}
  {2009})}\BibitemShut {NoStop}%
\bibitem [{\citenamefont {de'Medici}\ \emph {et~al.}(2005)\citenamefont
  {de'Medici}, \citenamefont {Georges},\ and\ \citenamefont
  {Biermann}}]{de'medici:205124}%
  \BibitemOpen
  \bibfield  {author} {\bibinfo {author} {\bibfnamefont {L.}~\bibnamefont
  {de'Medici}}, \bibinfo {author} {\bibfnamefont {A.}~\bibnamefont {Georges}},
  \ and\ \bibinfo {author} {\bibfnamefont {S.}~\bibnamefont {Biermann}},\
  }\href@noop {} {\bibfield  {journal} {\bibinfo  {journal} {Phys. Rev. B}\
  }\textbf {\bibinfo {volume} {72}},\ \bibinfo {pages} {205124} (\bibinfo
  {year} {2005})}\BibitemShut {NoStop}%
\bibitem [{\citenamefont {Kita}\ \emph {et~al.}(2011)\citenamefont {Kita},
  \citenamefont {Ohashi},\ and\ \citenamefont {Kawakami}}]{kita:195130}%
  \BibitemOpen
  \bibfield  {author} {\bibinfo {author} {\bibfnamefont {T.}~\bibnamefont
  {Kita}}, \bibinfo {author} {\bibfnamefont {T.}~\bibnamefont {Ohashi}}, \ and\
  \bibinfo {author} {\bibfnamefont {N.}~\bibnamefont {Kawakami}},\ }\href@noop
  {} {\bibfield  {journal} {\bibinfo  {journal} {Phys. Rev. B}\ }\textbf
  {\bibinfo {volume} {84}},\ \bibinfo {pages} {195130} (\bibinfo {year}
  {2011})}\BibitemShut {NoStop}%
\bibitem [{\citenamefont {Werner}\ and\ \citenamefont
  {Millis}(2006)}]{werner:155107}%
  \BibitemOpen
  \bibfield  {author} {\bibinfo {author} {\bibfnamefont {P.}~\bibnamefont
  {Werner}}\ and\ \bibinfo {author} {\bibfnamefont {A.~J.}\ \bibnamefont
  {Millis}},\ }\href@noop {} {\bibfield  {journal} {\bibinfo  {journal} {Phys.
  Rev. B}\ }\textbf {\bibinfo {volume} {74}},\ \bibinfo {pages} {155107}
  (\bibinfo {year} {2006})}\BibitemShut {NoStop}%
\bibitem [{\citenamefont {Werner}\ \emph {et~al.}(2006)\citenamefont {Werner},
  \citenamefont {Comanac}, \citenamefont {de' Medici}, \citenamefont {Troyer},\
  and\ \citenamefont {Millis}}]{werner:076405}%
  \BibitemOpen
  \bibfield  {author} {\bibinfo {author} {\bibfnamefont {P.}~\bibnamefont
  {Werner}}, \bibinfo {author} {\bibfnamefont {A.}~\bibnamefont {Comanac}},
  \bibinfo {author} {\bibfnamefont {L.}~\bibnamefont {de' Medici}}, \bibinfo
  {author} {\bibfnamefont {M.}~\bibnamefont {Troyer}}, \ and\ \bibinfo {author}
  {\bibfnamefont {A.~J.}\ \bibnamefont {Millis}},\ }\href@noop {} {\bibfield
  {journal} {\bibinfo  {journal} {Phys. Rev. Lett.}\ }\textbf {\bibinfo
  {volume} {97}},\ \bibinfo {pages} {076405} (\bibinfo {year}
  {2006})}\BibitemShut {NoStop}%
\bibitem [{\citenamefont {Gull}\ \emph
  {et~al.}(2011{\natexlab{a}})\citenamefont {Gull}, \citenamefont {Millis},
  \citenamefont {Lichtenstein}, \citenamefont {Rubtsov}, \citenamefont
  {Troyer},\ and\ \citenamefont {Werner}}]{gull:349}%
  \BibitemOpen
  \bibfield  {author} {\bibinfo {author} {\bibfnamefont {E.}~\bibnamefont
  {Gull}}, \bibinfo {author} {\bibfnamefont {A.~J.}\ \bibnamefont {Millis}},
  \bibinfo {author} {\bibfnamefont {A.~I.}\ \bibnamefont {Lichtenstein}},
  \bibinfo {author} {\bibfnamefont {A.~N.}\ \bibnamefont {Rubtsov}}, \bibinfo
  {author} {\bibfnamefont {M.}~\bibnamefont {Troyer}}, \ and\ \bibinfo {author}
  {\bibfnamefont {P.}~\bibnamefont {Werner}},\ }\href@noop {} {\bibfield
  {journal} {\bibinfo  {journal} {Rev. Mod. Phys.}\ }\textbf {\bibinfo {volume}
  {83}},\ \bibinfo {pages} {349} (\bibinfo {year}
  {2011}{\natexlab{a}})}\BibitemShut {NoStop}%
\bibitem [{\citenamefont {Gull}\ \emph
  {et~al.}(2011{\natexlab{b}})\citenamefont {Gull}, \citenamefont {Werner},
  \citenamefont {Fuchs}, \citenamefont {Surer}, \citenamefont {Pruschke},\ and\
  \citenamefont {Troyer}}]{gull:20111078}%
  \BibitemOpen
  \bibfield  {author} {\bibinfo {author} {\bibfnamefont {E.}~\bibnamefont
  {Gull}}, \bibinfo {author} {\bibfnamefont {P.}~\bibnamefont {Werner}},
  \bibinfo {author} {\bibfnamefont {S.}~\bibnamefont {Fuchs}}, \bibinfo
  {author} {\bibfnamefont {B.}~\bibnamefont {Surer}}, \bibinfo {author}
  {\bibfnamefont {T.}~\bibnamefont {Pruschke}}, \ and\ \bibinfo {author}
  {\bibfnamefont {M.}~\bibnamefont {Troyer}},\ }\href@noop {} {\bibfield
  {journal} {\bibinfo  {journal} {Comput. Phys. Comm.}\ }\textbf {\bibinfo
  {volume} {182}},\ \bibinfo {pages} {1078} (\bibinfo {year}
  {2011}{\natexlab{b}})}\BibitemShut {NoStop}%
\bibitem [{\citenamefont {B\"unemann}\ \emph {et~al.}(1998)\citenamefont
  {B\"unemann}, \citenamefont {Weber},\ and\ \citenamefont
  {Gebhard}}]{weber:6896}%
  \BibitemOpen
  \bibfield  {author} {\bibinfo {author} {\bibfnamefont {J.}~\bibnamefont
  {B\"unemann}}, \bibinfo {author} {\bibfnamefont {W.}~\bibnamefont {Weber}}, \
  and\ \bibinfo {author} {\bibfnamefont {F.}~\bibnamefont {Gebhard}},\
  }\href@noop {} {\bibfield  {journal} {\bibinfo  {journal} {Phys. Rev. B}\
  }\textbf {\bibinfo {volume} {57}},\ \bibinfo {pages} {6896} (\bibinfo {year}
  {1998})}\BibitemShut {NoStop}%
\bibitem [{\citenamefont {B\"unemann}\ \emph {et~al.}(2008)\citenamefont
  {B\"unemann}, \citenamefont {Gebhard}, \citenamefont {Ohm}, \citenamefont
  {Weiser},\ and\ \citenamefont {Weber}}]{bunemann:236404}%
  \BibitemOpen
  \bibfield  {author} {\bibinfo {author} {\bibfnamefont {J.}~\bibnamefont
  {B\"unemann}}, \bibinfo {author} {\bibfnamefont {F.}~\bibnamefont {Gebhard}},
  \bibinfo {author} {\bibfnamefont {T.}~\bibnamefont {Ohm}}, \bibinfo {author}
  {\bibfnamefont {S.}~\bibnamefont {Weiser}}, \ and\ \bibinfo {author}
  {\bibfnamefont {W.}~\bibnamefont {Weber}},\ }\href@noop {} {\bibfield
  {journal} {\bibinfo  {journal} {Phys. Rev. Lett.}\ }\textbf {\bibinfo
  {volume} {101}},\ \bibinfo {pages} {236404} (\bibinfo {year}
  {2008})}\BibitemShut {NoStop}%
\bibitem [{\citenamefont {Deng}\ \emph {et~al.}(2009)\citenamefont {Deng},
  \citenamefont {Wang}, \citenamefont {Dai},\ and\ \citenamefont
  {Fang}}]{deng:075114}%
  \BibitemOpen
  \bibfield  {author} {\bibinfo {author} {\bibfnamefont {X.}~\bibnamefont
  {Deng}}, \bibinfo {author} {\bibfnamefont {L.}~\bibnamefont {Wang}}, \bibinfo
  {author} {\bibfnamefont {X.}~\bibnamefont {Dai}}, \ and\ \bibinfo {author}
  {\bibfnamefont {Z.}~\bibnamefont {Fang}},\ }\href@noop {} {\bibfield
  {journal} {\bibinfo  {journal} {Phys. Rev. B}\ }\textbf {\bibinfo {volume}
  {79}},\ \bibinfo {pages} {075114} (\bibinfo {year} {2009})}\BibitemShut
  {NoStop}%
\bibitem [{\citenamefont {Lanat\`{a}}\ \emph {et~al.}(2012)\citenamefont
  {Lanat\`{a}}, \citenamefont {Strand}, \citenamefont {Dai},\ and\
  \citenamefont {Hellsing}}]{lanata:1108.0180}%
  \BibitemOpen
  \bibfield  {author} {\bibinfo {author} {\bibfnamefont {N.}~\bibnamefont
  {Lanat\`{a}}}, \bibinfo {author} {\bibfnamefont {H.~U.~R.}\ \bibnamefont
  {Strand}}, \bibinfo {author} {\bibfnamefont {X.}~\bibnamefont {Dai}}, \ and\
  \bibinfo {author} {\bibfnamefont {B.}~\bibnamefont {Hellsing}},\ }\href@noop
  {} {\bibfield  {journal} {\bibinfo  {journal} {Phys. Rev. B}\ }\textbf
  {\bibinfo {volume} {85}},\ \bibinfo {pages} {035133} (\bibinfo {year}
  {2012})}\BibitemShut {NoStop}%
\bibitem [{\citenamefont {de' Medici}\ \emph {et~al.}(2011)\citenamefont {de'
  Medici}, \citenamefont {Mravlje},\ and\ \citenamefont
  {Georges}}]{medici:256401}%
  \BibitemOpen
  \bibfield  {author} {\bibinfo {author} {\bibfnamefont {L.}~\bibnamefont {de'
  Medici}}, \bibinfo {author} {\bibfnamefont {J.}~\bibnamefont {Mravlje}}, \
  and\ \bibinfo {author} {\bibfnamefont {A.}~\bibnamefont {Georges}},\
  }\href@noop {} {\bibfield  {journal} {\bibinfo  {journal} {Phys. Rev. Lett.}\
  }\textbf {\bibinfo {volume} {107}},\ \bibinfo {pages} {256401} (\bibinfo
  {year} {2011})}\BibitemShut {NoStop}%
\end{thebibliography}%

\end{document}